\documentstyle[aps,preprint]{revtex}
\begin{document}
\draft
\title{REENTRANT WETTING TRANSITION OF A ROUGH WALL}
\author{Gilberto Giugliarelli}
\address{Dipartimento di Fisica and INFN(Gruppo Collegato di Udine, Sezione di
Trieste), Universit\'a degli Studi di Udine, I--33100
Udine, ITALY}
\author{Attilio L. Stella}
\address{Istituto Nazionale di Fisica della Materia -- Dipartimento di Fisica e
Istituto Nazionale di Fisica Nucleare (Sezione di Padova), Universit\'a di 
Padova, I--35131 Padova, ITALY}

\maketitle
\begin{abstract}
A $2D$ model describing depinning of an interface from a rough, self--affine
substrate, is studied by transfer matrix methods. The phase diagram is
determined for several values of the roughness exponent, $\zeta_S$, of the
attractive wall. For all $\zeta_S>0$ the following scenario is observed. In
first place, in contrast to the case of a flat wall  ($\zeta_S=0$), for wall
attraction energies between zero and a  $\zeta_S$--dependent positive value,
the substrate is always wet.  Furthermore, in a small range of attraction
energies, a dewetting transition  first occurs as $T$ increases, followed by a
wetting one. This unusual reentrance phenomenon  seems to be a peculiar feature
of self--affine roughness, and does not occur, e. g., for periodically
corrugated substrates.
\end{abstract}

\section{Introduction}

Wetting phenomena occur when, e. g., a layer of liquid phase coexisting with 
its vapor grows macroscopically over an attractive solid substrate
\cite{dietrich}. While wetting in pure systems is relatively well understood by
now,  the effects of disorder in interfacial phenomena
\cite{forgacs,huse,kardar&kardar,lipowski} pose many challenging issues and are
the object of active research. Of particular interest is the effect of
geometric surface disorder (roughness) on the location and the nature of the
wetting transition. An extensively studied type of roughness, also in view of
its experimental realizability \cite{pfeifer},  is that occurring when the
average height fluctuation in a sample of longitudinal linear size $L$, $w_L$,
scales like $w_L\sim L^{\zeta_S}$. This self--affine scaling disorder is
globally characterized by the roughness exponent $\zeta_S$, which can be
expected as the only relevant substrate parameter possibly affecting universal
features of the wetting transition. 

The relevance of $\zeta_S$ for both complete and critical wetting transitions
has been extensively studied in the last years \cite{li&kardar,giugliarelli1}.
As far as critical wetting is concerned, a recent study by the present authors
\cite{giugliarelli2} has put in evidence the fact that self--affine roughness
can produce a change from continuous to first--order transition in systems with
short--range forces, when the roughness of the substrate, $\zeta_S$, exceeds
the intrinsic roughness of the interface in the bulk, $\zeta_0$. On the other
hand, for $\zeta_S<\zeta_0$, no change is expected in the nature and
universality class of the continuous wetting transition, with respect to that
on flat substrate. 

The possibility that substrate roughness drives a continuous wetting
transition  first--order has been subsequently discussed in the context of
Landau--type mean field approaches \cite{parry}. 

A possible modification of the nature of the wetting transition is not the only
effect of self--affine substrate roughness on wetting. As we show here,
roughness produces modifications of the phase diagram of the interface, which
can be quite dramatic and important in experiments and applications. 

In the present paper we study systematically the wetting phase diagram of a
generalization \cite{giugliarelli2} of the Chui--Weeks model \cite{chui} with a
rough attractive boundary in 2$D$. By this study we produce evidence of some 
remarkable and definitely unusual features of the phase diagram, which, at a
qualitative level, should be considered as generic for wetting on self--affine
rough substrates. The most notable feature of the phase diagram is a reentrance
phenomenon for wetting in the whole range of roughnesses ($0<\zeta_S<1$). This
reentrance, which implies a dewetting followed by a wetting transition as the
temperature is raised, occurs both in regimes when the transitions (both
dewetting and wetting) are critical, and when they are first--order. 

\vskip8.0truecm
FIG. 1. Example of rough substrate wall (continuous path) and interface
configuration (dotted path).
\bigskip
\includegraphics{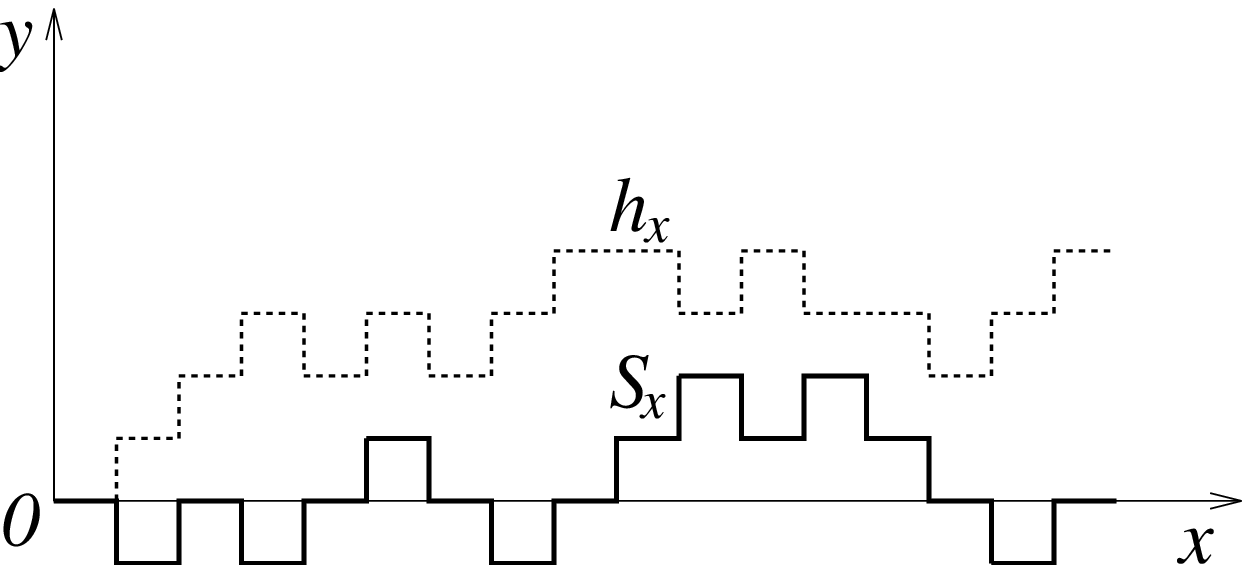}

This paper is organized as follows. In the next section we introduce the model
and describe the transfer matrix method we apply. In section 3 we dicuss how
the phase diagram is calculated and illustrate in detail the main results. The
last section is devoted to further general considerations and to conclusions.

\section{Model and Transfer Matrix}

We consider here the same generalization of the Chui--Weeks interfacial model
discussed in ref. \cite{giugliarelli2}.  Let us denote by $x$ and $y$ the
integer coordinates of points on a square lattice. The interface configurations
are self--avoiding paths (partially) directed in the $x$ direction (see Fig.
1).  Each interface step parallel to the $x$ axis is located by giving the
ordinate $y=h_x$ of its left--hand extremity. In this way the configuration is
determined by giving $h_x$, $\forall x$. We impose the following extra
restrictions on the set $\{  h_x\}$. First of all $h_x\geq S_x$ where $\{S_x\}$
represents the directed profile of the substrate wall. $\{S_x\}$ are sets of
integer ordinates determining the wall configuration with the same conventions
applied to $\{h_x\}$.  An extra constraint on possible interface configurations
is given by $h_{x+1}-h_x=0,\pm 1$, while for $S_x$ we choose $S_{x+1}-S_x=\pm
1$. We impose such constraints for computational convenience: their removal or
modification would not change the main qualitative features of the phase 
diagram.

The sets $\{S_x\}$ are generated by a random sampling procedure inspired by
Mandelbrot\cite{mandelbrot} and described in Appendix. This procedure produces
directed paths in 2$D$ obeying the restrictions described above and satisfying
the scaling relation 
\begin{equation}
\overline{|S_{x+\Delta x}-S_x|}\sim |\Delta x|^{\zeta_S}
\label{eq1}
\end{equation}
In the last equation the bar indicates statistical average with respect to the
above mentioned sample of $\{S_x\}$. Averages over wall configurations, i. e.
over different $\{S_x\}$, will be considered quenched and denoted by overbars
below.

The interface Hamiltonian is of the form
\begin{equation}
{\cal H}=\sum_x^X [\varepsilon (1+|z_x-z_{x-1}+s_x|)-u\delta_{z_x,0}]
\label{eq2} 
\end{equation}
with $z_x=h_x-S_x$ and $s_x=S_x-S_{x-1}$. In eq.(\ref{eq2}) $\varepsilon$ 
($\varepsilon >0$) is the energy cost of any interface step and $-u$ ($u>0$) is
the energy gain of an interface contact with the attracting wall. The sum in
(\ref{eq2}) is performed up to $X$, which represents the  length of the
interface projection on the $x$ axis. It should be noted that  in our model
only the horizontal steps of the interface paths in contact with  the wall are
prized by energies $-u$. This choice is again not mandatory. Different
conventions would not change the basic qualitative  results. 

At a finite temperature $T$ the fugacities $\omega=e^{-\varepsilon/T}$ and $k =
e^{u/T}$ are associated with each (horizontal or vertical) step of the path,
and to each horizontal step on the wall, respectively. Consequently, given a
wall profile, the partition function associated to all the interface
configurations covering a distance $X$ can be written in the form
\begin{equation}
{\cal Z}_X=\sum_{\{z_x\}} e^{-{1\over T} \left[\varepsilon \sum_x^X 
(1+|z_x-z_{x-1}+s_x|)-u\sum_x^X \delta_{z_x,0}\right] }=\omega^X
\sum_{\{z_x\}}
\omega^{n_\perp} k^{n_c}
\label{Z1}
\end{equation}
where the sum is done over the ensemble of all the directed  paths (determined
here by $\{z_x\}$) compatible with the choosen wall. $n_\perp$ and $n_c$ are
the number of vertical steps of the interfacial path and the number of
horizontal  steps on the wall, respectively. Note that the total length of a
path is given by $L=X+n_\perp$. 

The partition function (\ref{Z1}) can be more conveniently expressed in  the
form
\begin{equation}
{\cal Z}_X=\omega^X \sum_{l,z} \left(\prod_{x=1}^X {\bf
T}_{s_x}\right)_{l,z}
\phi_0(z)
\label{Z2}
\end{equation}
where the ${\bf T}_{s_x}$ are transfer matrices defined as
\begin{equation} 
({\bf T}_s)_{m,n}=[\delta_{m,n-s}+\omega
(\delta_{m,n-s-1}+\delta_{m,n-s+1})]k^{\delta_{m,0}}
\label{Tmatrix}
\end{equation}
with $m,n\geq 0$. The function $\phi_0$ in eq. (\ref{Z2}) can be used to
enforce particular initial conditions for the interface paths; if we choose
paths with an extremity on the wall, we put  $\phi_0(z)=\delta_{z,0}$.

With the above definitions, a wall profile corresponds to a particular sequence
of factors ${\bf T}_{s_x}$ in the product of eq.(\ref{Z2}). Correspondingly,
for asymptotically large systems, the partition function ${\cal Z}_X$ can be
expressed in terms of the largest Lyapunov eigenvalue $\lambda_{max}$ as ${\cal
Z}_X \sim (\omega\lambda_{max})^X$, for $X\to\infty$, where \cite{crisanti}
\begin{equation}
\lambda_{max}=\lim_{X\to\infty} \left[ || (\prod_{x=1}^X {\bf T}_{s_x})
{\vec \phi}_0 ||/ ||{\vec \phi}_0|| \right]^{1\over X}
\label{lambdamax1}
\end{equation}
with $||{\vec \phi}||=\sum_z \phi(z)$.\footnote{Note that our definition of
norm is legitimate because $\phi_0(z)$ and the elements of the ${\bf T}_{s_x}$'s
matrices are $\geq 0$.} 

For walls with some periodic geometry with period $X_p$ the matrix product in
(\ref{lambdamax1}) can be written as  $[{\cal T}_{X_p}]^{X/X_p}$ ($X$ assumed 
an integer multiple of $X_p$) with ${\cal T}_{X_p}=\prod_{x=1}^{X_p} {\bf
T}_{s_x}$. In this case the calculation of $\lambda_{max}$ is equivalent to the
calculation of the largest eigenvalue, $\Lambda_{X_p}$, of the matrix ${\cal
T}_{X_p}$. Consequently, $\lambda_{max}= [\Lambda_{X_p}]^{1/X_p}$.

In numerical calculations it is useful to introduce the normalized vectors 
${\vec \psi}_x$ defined by the recursion relation
\begin{equation}
{\vec \psi}_x= {1\over {n_x}} {\bf T}_{s_x} {\vec \psi}_{x-1}
\label{psi}
\end{equation}
with $n_x=||{\bf T}_{s_x} {\vec \psi}_{x-1}||$ and  ${\vec \psi}_0\equiv {\vec
\phi}_0$. It is straightforward to see that the  $z$--th  component of the
vector ${\vec \psi}_x$  corresponds to the probability that the path at $x$ is
at a distance $z$ from the wall \cite{forgacs,giugliarelli1}. This suggests
our  definition of $||{\vec \psi}_x||$. 
For a given wall profile, the above  definitions allow to express  the Lyapunov
eigenvalue (\ref{lambdamax1}) as
\begin{equation}
\lambda_{max}=\lim_{X\to\infty}\left[\prod_{x=1}^X n_x\right]^{1\over X}=
 \exp\left( \lim_{X\to\infty} {1\over X} \sum_{x=1}^X \ln n_x \right)
\label{lambdamax2}
\end{equation}
The quenched, dimensionless free energy density  given by    $\lim_{X\to\infty}
-\overline{\ln {\cal Z}_X}/X$  can be written in the following form 
\begin{equation}
\overline{f}=-\overline{\ln 
\omega\lambda_{max}}=-\ln\omega-\lim_{X\to\infty} 
{1\over X}\overline{\sum_{x=1}^X \ln n_x}. 
\label{f}
\end{equation}
If $X$ is chosen large enough, the average over quenched wall disorder  for the
second term on the right hand side requires only a rather limited sample of
wall configurations. This is due to a self--averaging property of $f$, which
clearly manifests itself in the numerical results.

\section{Depinning transition and Phase diagram}

In the discussion below we implicitly make $u$ and $T$ dimensionless by
dividing them by $\varepsilon$. The wetting transition occurs because, e. g.,
at a given $T$, the interface can be bound to the wall of the substrate only
for sufficiently high values of $u$. In the case of ordered flat walls, i. e.
$\{S_x=constant \ \forall x\}$, the value of $u$ above which the interface is
pinned can be easily calculated \cite{abraham,chui,forgacs} to be 
\begin{equation}
u_c(T)=T\ln\left[(1+2\exp(-1/T))/(1+\exp(-1/T))\right]
\label{uc_flat}
\end{equation}
with $u_c(0)=0$. We denote by $P_0$ the average fraction of interface
horizontal steps on the  wall, 
\begin{equation}
P_0=\lim_{X\to\infty} \langle n_c\rangle /X, 
\label{P0}
\end{equation}
with brackets indicating canonical thermal average.  One can check that $P_0$
vanishes continuously and linearly when the line $u=u_c(T)$ is approached from
above.

When dealing with our random substrates the calculation of  $P_0$ or $f$ for
each $\{S_x\}$ can not be done exactly in a semi--infinite geometry, and
truncations must be made. To minimize effects of the finite size  of the
transfer matrices, in our calculations we only considered matrix sizes  much
larger than the mean square perpendicular width of the self--affine walls. This
width can be defined as $\Delta  S_\perp=(\overline{ S^2} -{\overline{
S}}^2)^{1/2}\sim  X^{\zeta_S}$, with $\overline{S^2}=(1/X)\sum_x S_x^2$ and 
$\overline{S}=(1/X)\sum_x S_x$. In practice we used transfer matrices as large
as $10^4\times 10^4$ in the roughest case, corresponding to $\zeta_S=\ln 12/\ln
32\simeq  0.717$. With this roughness $X=10^5$ was reached. 

In addition one has to average over different $\{S_x\}$ in order to get
$\overline{P_0}$ and $\overline{f}$. In practice we could sample at most 10 or
15 different $\{S_x\}$ in the most favorable cases, due to the large $X$'s
needed to  extract precisely $P_0$. Fortunately, as mentioned above, our large
$X$ values, of the  order of $10^5$, lead to a very high degree of
self--averaging in quantities like $f$ and $P_0$. 

At fixed $T$, as the transition is approached from above (viz. $u>u_c(T)$) the
interface free energy density (\ref{f}) is negative and increasing, with
decreasing $u$; at $u=u_c(T)$ it matches the bulk interface  free energy
density $f_{bulk}=-\ln\omega(1+2\omega)$. Thus, it is useful to consider the
interface excess free energy density, $\Delta f=\overline{f}-f_{bulk}$, which,
by means of eq. (\ref{f}), can be expressed as
\begin{equation}
\Delta f= -\overline{\ln \lambda_{max}} +\ln
(1+2\omega)=-\overline{\ln\lambda'_{max}}
\label{Deltaf}
\end{equation}
This last equation defines $\lambda'_{max}$  as the Lyapunov eigenvalue of 
transfer matrices defined as ${\bf T}'_s=[1/(1+2\omega)]{\bf T}_s$, which were
directly used in our calculations.    
 
Thus  the position of the depinning transition corresponds to the vanishing of
$\Delta f$.  Also the interface contact probability can be evaluated using
$\lambda'_{max}$, taking into account that
\begin{equation}
\overline{P_0}=k{{\partial\Delta f}\over {\partial k}}=
\left({k\over {\lambda'_{max}}}\right)
\overline{{{\partial\lambda'_{max}}\over {\partial k}}}=
\left({T\over{\lambda'_{max}}}\right) 
\overline{{{\partial\lambda'_{max}}\over {\partial u}}}
\label{P0_deriv}
\end{equation}

\vskip14.0truecm
FIG. 2. Interface phase diagram for rough self--affine walls in the $T$--$u$
plane ($\varepsilon=1$ is assumed).  The curves $u=u_c(T)$ are shown for five
different values of the roughness exponent $\zeta_S$.  The light 
continuous line corresponds to $u=u_c(T)$ for a flat wall as given by eq. 
(\ref{uc_flat}).
\bigskip
\includegraphics{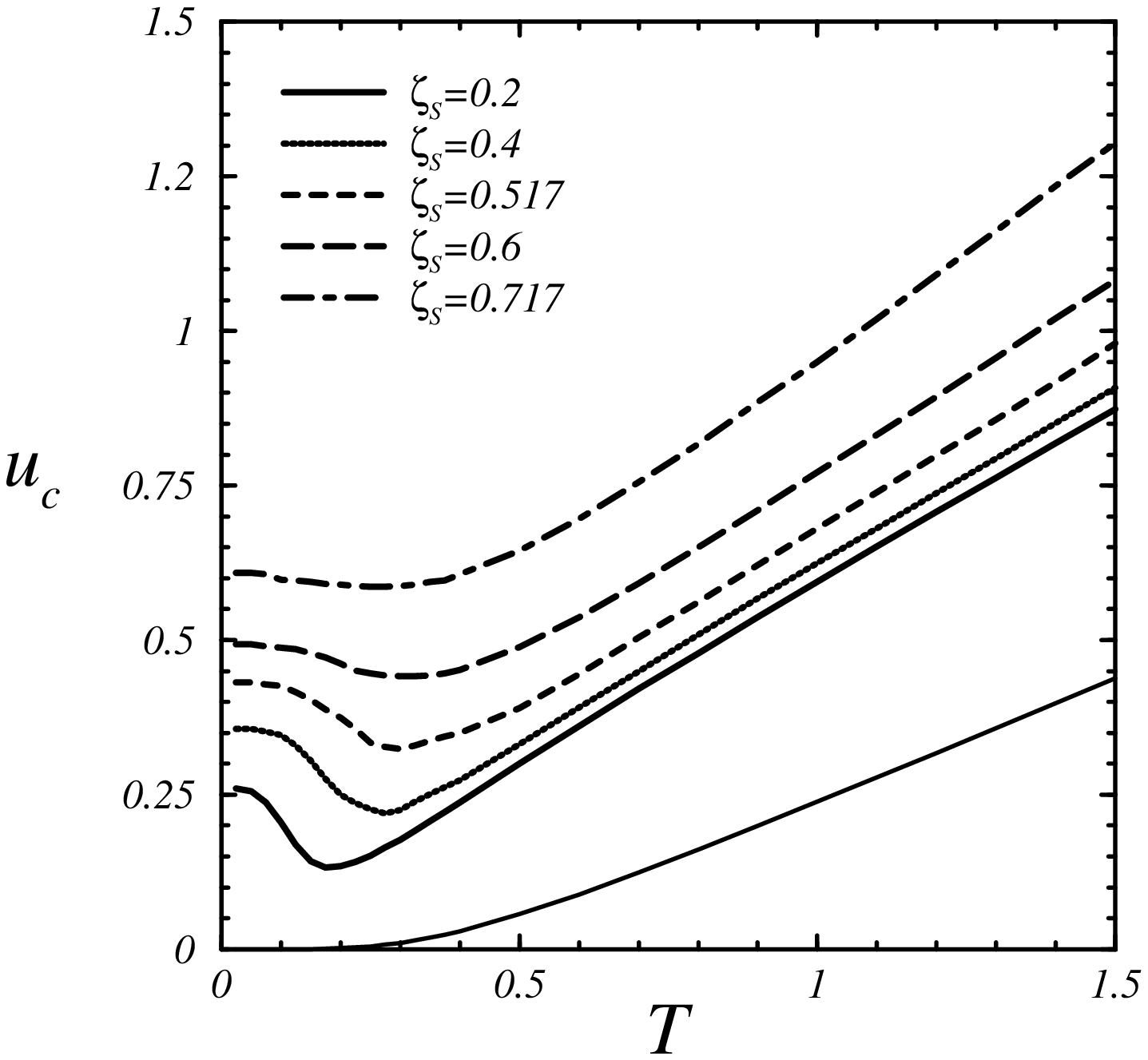}

The calculation of $\overline{P_0}$ offers an alternative way of locating
dewetting or wetting transitions, by identifying the conditions under which the
quantity first becomes zero. A numerical study of how  $\overline{P_0}$
approaches zero can also provide informations on whether these transitions are
continuous or discontinuous\cite{giugliarelli2}. We used this criterion,
together with an analysis of the way in which $\Delta f$ vanishes to clarify
the first or second order nature of the transitions.

Fig. 2 summarizes the results we have obtained by a systematic calculation of
$\overline f$ as function of $T$ and $u$, for 5 different values of the 
$\zeta_S$. The curves in the figure represent the behavior of $u_c$ versus $T$.
$u_c$ was determined numerically as the value of $u$ below which $|\Delta
f|\leq 0.0001$ as $u$ was changed in steps of $0.001$. By  using the definition
of $\omega$ and $k$ the $u=u_c(T)$ curves can be converted into the curves
$k=k_c(\omega)$. Note that $u_c(0)\neq 0$ implies a power--law divergence of
$k_c(\omega)$ for $\omega \to 0$: $k_c\simeq \omega^{-u_c(0)}$.

Looking at the curves for rough walls in Fig. 2, we note two main  differences
from the flat case: 
\begin{itemize}
\item [a)] It is not possible to pin an interface to a rough wall with a
vanishing contact energy. As $T\to 0$ the minimal contact energy to  pin an
interface, $u_c(0)$, is finite and increases as the wall roughness, i. e.
$\zeta_S$, increases.
\item [b)] For each $\zeta_S$, there is a  temperature,  $T_R(\zeta_S)$, below
which $u_c$ is a decreasing function of $T$.  Note also that for all
$\zeta_S<1$ $u_c(T)$ approaches $u_c(0)$ with zero slope.
\end{itemize}

A surprising consequence of b) is that if, as in an experiment, we monitor 
interface behavior at fixed $u$, by varying $T$, we have to distinguish among
three kinds of regimes:
\begin{itemize}
\item [I)] For $u<u_c(T_R)$ interface pinning is impossible, no matter
how low $T$ is. The substrate is wet at all $T$, and no transitions take
place.
\item [II)] for $u>u_c(0)$ as the temperature is increased the interface
passes  from a pinned to a depinned state at some $T_W$. Thus, the 
substrate is partially wet for $T<T_W$ and wet for $T>T_W$, as for smooth walls.
\item [III)] for $u_c(T_R)<u<u_c(0)$, as the temperature is increased, the
interface undergoes two transitions: 1) at some temperature  $T_D<T_R$ we find
an unexpected dewetting: the interface which is  depinned for $T<T_D$ becomes
pinned at $T=T_D$;  2) at  some $T_W>T_R$ a  more usual depinning transition
follows. Thus, the substrate is wet at very low $T$, then it dewets, and
eventually it wets again at high $T$. 
\end{itemize}

Unfortunately, a detailed study of the behavior of $T_R$ for $\zeta_S$
approaching zero is not feasible due to the necessity of generating extremely
long walls in order to distinguish, e. g., $\zeta_S=0.1$ from $\zeta_S=0$.
However, our results suggest rather clearly that $T_R$ approaches zero for both
$\zeta_S\to 0$ and $\zeta_S\to 1$. Thus, in these two limits the reentrance
desappears. 

Another interesting aspect of the phase diagram is that connected to the nature
of the transitions involved. The continuous or discontinuous character of the
wetting transitions upon varying $\zeta_S$ was already discussed  in ref.
\cite{giugliarelli2} by analyzing the way in which $\overline{P_0}$ approaches
zero for $u\to u_c(T)$. There we found that when $\zeta_S\lesssim 1/2$ the
transition remains second--order and most likely belongs to the same
universality class as the flat case, i. e. $\overline{P_0}\sim (u-u_c)^\psi$,
with $\psi=1$. When $\zeta_S$ exceeds 1/2 there is clear evidence that
depinning occurs discontinuously. $\zeta_0=1/2$ is the roughness exponent of
the interface in the bulk and it makes sense that this is the precise border
value of $\zeta_S$ separating the  two regimes.

A natural question is whether, in the range $u_c(T_R)<u<u_c(0)$, the dewetting
transition occurring at low temperature is of the same kind as its wetting
counterpart at higher $T$.  Following the lines of ref. \cite{giugliarelli2} 
we made a systematic study of the dewetting transition for two $\zeta_S$
values, respectively below and above $\zeta_S=1/2$. In the first case
($\zeta_S=0.4$)  we found evidence of a continuous dewetting, while in the
latter ($\zeta_S=0.6$)  it appeared discontinuous. Thus, in spite of the fact
that the dewetting transition occurs only for some $u$ ranges, and at much
lower temperatures, it seems that its character could be the same as that of
the corresponding wetting transition. Of course, a more precise definition of
the limits within which the above identity of transition orders applies would
require an extensive systematic exploration, which is beyond the scope of the
present work.

\section{Conclusions}

The results presented in the previous section are somewhat unexpected and are
worth discussing further.

\vskip8.0truecm
FIG. 3. Example of a periodically corrugated wall on a square lattice (heavy
line). The wall period is $X_p=10$. The light straight horizontal line
corresponds to the bound interface ground state at $T=0$ and $u \gtrsim 0$. 
\includegraphics{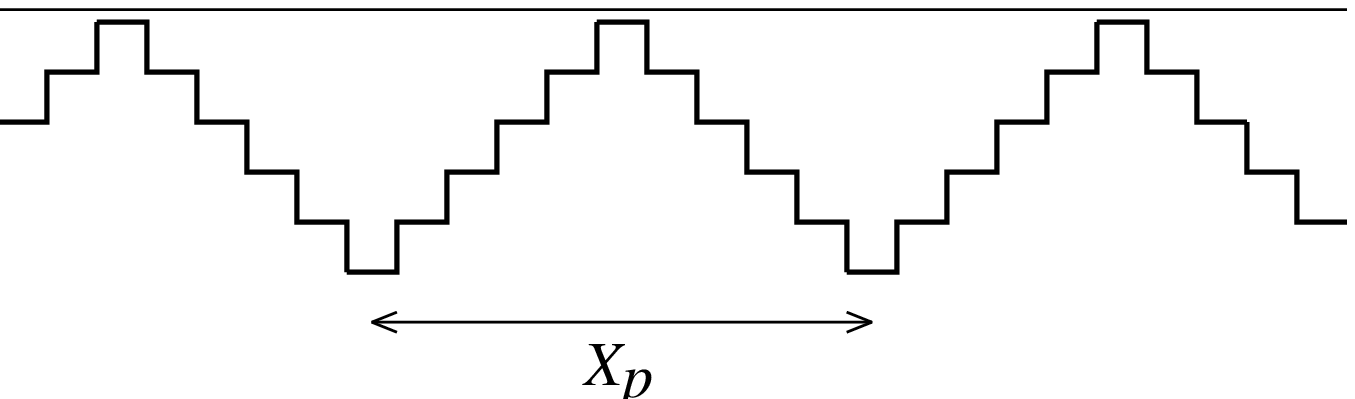}

The first important fact is that $u_c(0)>0$ for $\zeta_S>0$. At $T=0$, in order
to decide whether the interface is pinned or not, we need only to  compare the
ground state energy in the bulk with the lowest energy of a state in which the
interface is bound to the substrate. In the bulk the state of lowest possible
energy is clearly given by a straight  configuration ($n_\perp=0$). A bound
state will have an energy relative  to this unbound ground state equal to
$n_\perp-u n_c$ (assuming $\varepsilon=1$). $n_\perp$ and $n_c$ of course
depend on the wall configuration to which this bound state refers. Clearly
$u_c(0)$ is determined by the condition  under which this energy difference
between the two states vanishes: $u_c(0)=\lim_{X\to\infty} n_\perp /n_c$. The
fact that bound ground state  configurations satisfy this limit condition with
$u_c(0)>0$ is a nontrivial  property of self--affine substrates.  On a
periodically corrugated substrate with average horizontal orientation (like
that sketched in Fig. 3) this limit property would not be  satisfied. In that
case, for $u$ very close to zero, the bound  ground state configuration is  one
in which $n_\perp=0$, corresponding to a straight interface touching the
attractive tips of the periodically corrugated wall (see Fig. 3).  Thus, we
would have $n_\perp=0$ and $n_c\ne 0$, and, consequently, $u_c(0)=0$, like in a
flat case. We conclude that a remarkable property of self--affine substrates is
that they can support ground state interface configurations with
$\lim_{X\to\infty}n_\perp/n_c>0$.

Notice further that for a periodically corrugated substrate the reentrance 
phenomenon with dewetting preceeding wetting is not possible, since in that
case the curve $u=u_c(T)$ can not be decreasing in the  neighborhood of $T=0$.
We verified by explicit calculations that for a wall as in Fig. 3, $u_c(T)$ is
in fact never decreasing on the whole  $T$ axis. 

An even more remarkable property of the self--affine substrate, for which 
$u_c(0)>0$ is clearly a necessary but not sufficient condition, is the 
monotonically decreasing character of the curve $u=u_c(T)$ in the interval
$(0,T_R)$. This feature implies that, as soon as  $T>0$, an interface can be
more easily bound to the rough substrate. This clearly shows that there is a
very nontrivial energy--entropy interplay in the pinning mechanism when
self--affine roughness is involved.   

We also compared the nature of dewetting transitions to the corresponding
high--$T$ wetting ones. We got only preliminary results on this issue. These
results suggest the possibility that, once a first--order character prevails
for wetting, the same applies to dewetting as well. This would mean that
geometry alone is the crucial factor in determining the nature of transitions
on rough substrates.

The model calculations we presented here are of course limited to $2D$ and to
strictly short--range forces. An extension in $3D$ is computationally
unfeasible, and also the inclusion of long--range potentials would pose serious
additional difficulties in our calculations.  In $3D$ we expect that the main
features of the phase diagram would persist. Of course a major difference in
3$D$ would be the character of the transitions. In $3D$ the interface has
roughness $\zeta_0=0$ in the bulk \cite{forgacs}. Thus, following the
conclusions of ref. \cite{giugliarelli2},  we should  probably expect
first--order dewetting and wetting transitions for all $\zeta_S>0$.

Concerning the effect of long--range forces, which should certainly be included
in more realistic calculations to compare with experiments, we can only
conjecture that they would not modify the main result obtained here, i. e. the
reentrance. However, there is at least one instance, that of interfaces in
superconductors \cite{indekeu}, in which a short range  description is fully
adequate.
 
\acknowledgements

This work was partially supported by CNR within the CRAY Project of Statistical
Mechanics.  We are very grateful to T.L. Einstein for a critical reading of the
manuscript.

\appendix

\section*{Generating random self--affine paths in $2D$}

The procedure we describe here is a random version of a deterministic algorithm
by Mandelbrot \cite{mandelbrot}. Given two even integers, $p$  and $q$, with $p
< q$, the procedure allows us to construct iteratively a partially directed
path with a roughness exponent $\zeta_S= \ln p/\ln q$ and with $X=q^n$ after
$n$ iterations.

First we consider a set $\{ {\vec \alpha } \}$ of vectors with $q$ components.
For each vector $\vec \alpha$ in the set $(q+p)/2$ components, chosen at
random, are set equal to $1$, while the remaining $(q-p)/2$ are put equal to
$-1$. Once the set $\{ {\vec \alpha} \}$ contains a sufficiently large  number
of such elements, the construction of a wall profile proceeds iteratively. In
the first iteration we consider a vector ${\vec \beta}_1$ with $q$  components
and, after choosing an element $\vec \alpha$ at random in the set  $\{{\vec
\alpha} \}$, we set ${\vec \beta}_1={\vec \alpha}$.

In the second iteration we construct a vector ${\vec \beta}_2$ with $q^2$
components. Once chosen $q$ vectors ${\vec \alpha}_j$ ($j=1,2,\ldots,q$) at
random in $\{ {\vec \alpha} \}$,  the components of ${\vec \beta}_2$ are
determined according to the rule: 
\begin{equation}
\beta_2((j-1)q+i)=\beta_1(j)\alpha_j(i)
\end{equation}
with $i=1,2,\ldots,q$ and $\alpha_j(i)$ indicating the $i$--th component of the
vector ${\vec \alpha}_j$. 

The last equation can obviously be iterated for the construction of vectors
${\vec \beta}_n$ with $q^n$ components: 
\begin{equation}
\beta_n((j-1)q^{n-1}+i)=\beta_{n-1}(j)\alpha_j(i)
\end{equation}
with $j=1,2,\ldots,q^{n-1}$ and $i=1,2,\ldots,q$. 

At any iteration $n$ a directed path defined by the sequence of integers $S_x$,
as explained in section 2, can be obtained from ${\vec \beta}_n$ by
\begin{equation}
S_{x+1}=S_x+\beta_n(x)
\end{equation}
where we usually set $S_1=0$. Larger $n$ implies a large horizontal size
$X=q^n$ of the path. The accuracy of the self--affine average scaling of  the
paths, as detectable on the basis of eq.(1), increases with  increasing $n$ and
the number of walls considered. For $q$ ranging between 8 and 32 we have seen
that when the longitudinal length of the paths $X$ is of the order of $10^5$
the computed roughness exponents of a single generated path coincides with the
theoretical $\zeta_S=\ln p/\ln q$ within 1\%.

\end{document}